\documentclass[twocolumn,trackchanges]{aastex7}

\usepackage{textcomp}
\usepackage{xspace}
\usepackage{amsmath}
\usepackage{amssymb} 
\usepackage{graphicx}
\usepackage{float}
\usepackage{gensymb}
\usepackage{comment}
\usepackage{gensymb}
\usepackage{bm}

\newcommand {\ixpe}{\text{IXPE}\xspace}

\newcommand{\maxi}{\text{MAXI}\xspace}
\newcommand{\nustar}{\text{NuSTAR}\xspace}

\begin{document}

\title{Discovery of Strong Energy-Dependent X-ray Polarization in the Intermediate State of GS~1354$-$64}

\author[orcid=0000-0002-2381-4184,sname=Ravi,gname=Swati]{Swati Ravi}
\affiliation{MIT Kavli Institute for Astrophysics and Space Research, Massachusetts Institute of Technology \\ 77 Massachusetts Avenue, Cambridge, MA 02139, USA}
\email[show]{swatir@mit.edu}  

\author[orcid=0009-0001-4644-194X]{Lorenzo Marra}
\affiliation{INAF Istituto di Astrofisica e Planetologia Spaziali, Via del Fosso del Cavaliere 100, 00133 Roma, Italy}
\email{lorenzo.marra@inaf.it}

\author[orcid=0000-0002-5872-6061,gname=James F., sname='Steiner']{James F. Steiner} 
\affiliation{Center for Astrophysics \textbar\ Harvard \& Smithsonian, 60 Garden Street, Cambridge, MA 02138, USA}
\email{james.steiner@cfa.harvard.edu}

\author[0000-0003-4216-7936]{Guglielmo Mastroserio}
\affiliation{Scuola Universitaria Superiore IUSS Pavia, Palazzo del Broletto, piazza della Vittoria 15, I-27100 Pavia, Italy}
\email{guglielmo.mastroserio@iusspavia.it}

\author[0000-0002-0940-6563, gname=Mason, sname=Ng]{Mason Ng}
\affiliation{Department of Physics, McGill University, 3600 rue University, Montr\'{e}al, QC H3A 2T8, Canada}
\affiliation{Trottier Space Institute, McGill University, 3550 rue University, Montr\'{e}al, QC H3A 2A7, Canada}
\email{mason.ng@mcgill.ca} 

\author[orcid=0000-0002-8247-786X,sname='Neilsen']{Joey Neilsen}
\affiliation{Villanova University, Department of Physics, Villanova, PA 19003, USA}
\email{jneilsen@villanova.edu}

\author[orcid=0000-0002-6492-1293, sname=Marshall,gname=Herman]{Herman L. Marshall}
\affiliation{MIT Kavli Institute for Astrophysics and Space Research, Massachusetts Institute of Technology \\ 77 Massachusetts Avenue, Cambridge, MA 02139, USA}
\email{hermanm@mit.edu}

\author[0000-0002-6384-3027]{Fiamma Capitanio}
\affiliation{INAF Istituto di Astrofisica e Planetologia Spaziali, Via del Fosso del Cavaliere 100, 00133 Roma, Italy}
\email{fiamma.capitanio@inaf.it}

\author[orcid=0000-0001-5975-1026]{Sudeb Ranjan Datta}
\affiliation{Inter-University Centre for Astronomy and Astrophysics, Post Bag 4, Ganeshkhind, Pune - 411007, India}
\email{sudeb.datta@iucaa.in} 

\author[0000-0002-1532-4142]{Elise Egron}
\affiliation{INAF Osservatorio Astronomico di Cagliari, Via della Scienza 5,
09047 Selargius (CA), Italy}
\email{elise.egron@inaf.it}

\author[0000-0003-3828-2448]{Javier A. Garc\'ia}
\affiliation{X-ray Astrophysics Laboratory, NASA Goddard Space Flight Center, Greenbelt, MD 20771, USA}
\affiliation{Cahill Center for Astronomy \& Astrophysics, California Institute of Technology,
Pasadena, CA 91125, USA}
\email{javier.a.garciamartinez@nasa.gov}

\author[0000-0002-5311-9078]{Adam Ingram}
\affiliation{School of Mathematics, Statistics, and Physics, Newcastle University, Newcastle upon Tyne NE1 7RU, UK}
\email{Adam.Ingram@newcastle.ac.uk}

\author[0000-0002-3638-0637]{Philip Kaaret}
\affiliation{NASA Marshall Space Flight Center, Huntsville, AL 35812, USA}
\email{philip.kaaret@nasa.gov}

\author[0000-0001-8670-4575]{Ole~K\"onig}
\affiliation{Center for Astrophysics \textbar\ Harvard \& Smithsonian, 60 Garden Street, Cambridge, MA 02138, USA}
\email{ole.koenig@cfa.harvard.edu}

\author[orcid=0000-0003-2845-1009, sname=Liu,gname=Honghui]{Honghui Liu}
\affiliation{Institut f\"ur Astronomie und Astrophysik, Universit\"at T\"ubingen, Sand 1, D-72076 T\"ubingen, Germany}
\email{honghui.liu@uni-tuebingen.de}

\author[0000-0001-7374-843X]{Romana Miku\v{s}incov\'{a}}
\affiliation{INAF Istituto di Astrofisica e Planetologia Spaziali, Via del Fosso del Cavaliere 100, 00133 Roma, Italy}
\email{romana.mikusincova@inaf.it}

\author[orcid=0000-0002-9633-9193, sname=Nathan,gname=Edward]{Edward J. R. Nathan}
\affiliation{NASA Postdoctoral Program Fellow, X-ray Astrophysics Laboratory, NASA Goddard Space Flight Center, Greenbelt, MD 20771, USA}
\email{edward.nathan@nasa.gov}

\author[orcid=0000-0001-6061-3480,gname=Pierre-Olivier, sname=Petrucci]{P.-O. Petrucci}
\affiliation{Universit\'e Grenoble Alpes, CNRS, IPAG, 38000 Grenoble, France}
\email{pierre-olivier.petrucci@univ-grenoble-alpes.fr}

\author[0000-0001-5418-291X]{Jakub Podgorn\'y}
\affiliation{Astronomical Institute of the Czech Academy of Sciences, Bo\v{c}n\'i II 1401/1, 14100 Praha 4, Czech Republic}
\email{jakub.podgorny@asu.cas.cz}

\author[orcid=0000-0003-4472-1232]{Chiara Salvaggio}
\affiliation{INAF – Osservatorio Astronomico di Brera, Via E. Bianchi 46, 23807 Merate, (LC), Italy}
\email{chiara.salvaggio@inaf.it}

\author[0000-0003-2931-0742]{Ji\v{r}\'i~Svoboda}
\affiliation{Astronomical Institute of the Czech Academy of Sciences, Bo\v{c}n\'i II 1401/1, 14100 Praha 4, Czech Republic}
\email{jiri.svoboda@asu.cas.cz}  

\author[0000-0002-5767-7253]{Alexandra Veledina}
\affiliation{Department of Physics and Astronomy, 20014 University of Turku, Finland}
\affiliation{Nordita, KTH Royal Institute of Technology and Stockholm University, Hannes Alfv\'ens v\"ag 12, SE-10691 Stockholm, Sweden}
\email{alexandra.veledina@gmail.com}

\author[orcid=0000-0002-2268-9318,gname=Yuexin, sname=Zhang]{Yuexin Zhang} 
\affiliation{Center for Astrophysics \textbar\  Harvard \& Smithsonian, 60 Garden Street, Cambridge, MA 02138, USA}
\email{yuexin.zhang@cfa.harvard.edu} 

\begin{abstract}

We report the discovery of significant X-ray polarization from the dynamically confirmed black hole X-ray binary (BHXB) GS~1354$-$64 during its 2025--2026 outburst, obtained with the Imaging X-ray Polarimetry Explorer (\ixpe). The observation, obtained shortly after a bright X-ray flare, captures the source in an intermediate state following a stalled (failed) state transition. We discover significant 2--8~keV polarization at the $\sim$4$\%$ level with high statistical support---$14\sigma$ significance from frequentist analysis and log Bayes Factor $\Delta\mathrm{ln}(Z)=283\pm1$ in a Bayesian framework---measuring ${\rm PD}=4.0\pm0.2\%$ and ${\rm PA}=-1^\circ\pm2^\circ$ (90\% credible interval). The PD exhibits a statistically significant increasing trend with energy---the strongest such increase yet observed by \ixpe in a BHXB---going from $2.1\pm0.3\%$ in the 2--3~keV band to $11\pm3\%$ in the 6.5--7~keV band, while the PA appears stable across both energy and time to within statistical uncertainties. Timing analysis of the \ixpe data reveals a $\sim$5~Hz Type-C quasi-periodic oscillation. \ixpe+ \nustar spectropolarimetric modeling suggests that the data can be described by polarized thermal disk and Comptonized components with PAs differing by $\sim$90$^\circ$, or instead, our favored scenario with a dominant polarized Compton component whose PD increases across the \ixpe bandpass---the inferred component-level polarization levels are therefore model-dependent. In either picture, GS~1354$-$64 retains a strong coronal component during the transitional period observed by \ixpe. These results illustrate how X-ray polarimetry can provide a sensitive diagnostic of the accretion state and geometry in black hole X-ray binary accretion flows, exploring a liminal phase at the cusp of state transition.

\end{abstract}


\section{Introduction} 

Black hole X-ray binaries (BHXBs) exhibit distinct accretion states characterized by dramatic spectral and timing changes, generally interpreted as changes in the geometry and relative dominance of the accretion disk and hot Comptonizing corona \citep{2006ARA&A..44...49R, 2007A&ARv..15....1D}. State transitions occur on timescales typically from days to weeks and are accompanied by rapid evolution in luminosity, spectral hardness, and reflection features \citep{2004MNRAS.355.1105F, 2005Ap&SS.300..107H}. Yet despite decades of study, the evolution of the inner disk radius and coronal structure remains a topic of active debate.

Over the past four years, X-ray polarimetry with the Imaging X-ray Polarimetry Explorer \citep[\ixpe;][]{2021AJ....162..208S, 2022JATIS...8b6002W} has begun to directly probe accretion geometry in BHXBs, providing the first empirical constraints on how polarization properties evolve with spectral state (e.g., \citealt{2022Sci...378..650K, 2024ApJ...969L..30S, 2024A&A...686L..12P, 2024ApJ...968...76I}). Observations of Cygnus~X-1 in the soft state have revealed, for example, a modest polarization degree (PD) with a stable polarization angle (PA), consistent with disk-dominated emission viewed at moderate inclination \citep{2024ApJ...969L..30S}, while hard-state observations are dominated by the coronal contribution over \ixpe's bandpass \citep{2022Sci...378..650K} and favor an accretion geometry with the corona elongated in the plane of the disk. More recently, \ixpe captured the hard-to-soft transition of the BHXB Swift~J1727.8$-$1613 during its 2023 outburst \citep{2023ApJ...958L..16V,2024ApJ...966L..35S, 2024ApJ...968...76I, 2024A&A...686L..12P}, revealing a pronounced decrease in PD during the soft state, contrasted with a $\sim$3--4$\%$ PD in hard/intermediate states, tracking accretion disk contribution to the total X-ray flux. This offers direct evidence that X-ray polarization is sensitive to state-dependent changes in accretion geometry and emission mechanisms \citep{2024ApJ...966L..35S, 2024ApJ...968...76I, 2024A&A...686L..12P,2025A&A...701A.115K}. More generally, changes in observed X-ray polarization provides an independent probe of the geometry and orientation of the inner accretion flow to traditional spectroscopic methods \citep{2009ApJ...701.1175S, 2010ApJ...712..908S, 2011ApJ...731...75D}, even in the presence of strong winds (e.g., \citealt{2025A&A...694A.230N}).

GS~1354$-$64 (also known as BW~Cir) is a dynamically confirmed BHXB with a mass function of $f(M)\simeq5.7M_\odot$, indicating a black hole primary \citep{2004ApJ...613L.133C}. The system is thought to be viewed at relatively high inclination (e.g., $i\lesssim79^\circ$ reported in \citealt[]{2009ApJS..181..238C} and $i\approx75^\circ$ in \citealt[]{2016ApJ...826L..12E}) though this remains debated \citep{2023ApJ...951..145L, 2024ApJ...969...40D}. While it was previously placed at an extreme distance of $\sim20$–$25~\mathrm{kpc}$---among the most distant proposed Galactic BHXBs \citep{2009ApJS..181..238C}---\textit{Gaia} astrometry instead situates it much closer, at only $0.7^{+3.4}_{-0.3}~\mathrm{kpc}$ \citep{2019MNRAS.485.2642G}. 

Since its luminous 1987 discovery outburst \citep{1990ApJ...361..590K}, GS~1354$-$64 has undergone outbursts in 1997 \citep{2001MNRAS.323..517B}, 2015 \citep{2016MNRAS.460..942K, 2016MNRAS.459.4038S, 2017MNRAS.469..193P} and 2025--2026 \citep{2025ATel17563....1N}. Unlike the 1987 event, these later outbursts have been ``stalled'' or ``failed'' transitions \citep{2016MNRAS.460..942K, 2016MNRAS.459.4038S}, with the source remaining in the hard and hard-intermediate states without reaching a canonical soft state. Spectral studies of GS~1354$-$64 during previous outbursts have revealed strong Comptonization and, in some epochs, signatures of relativistic reflection from the inner accretion disk \citep{2001MNRAS.323..517B, 2016MNRAS.459.4038S, 2016MNRAS.460..942K}. 
However, fits for the black hole spin and inner-disk geometry have produced varied results.
In BHXBs, it is often instructive to compare the PA to the radio jet's position angle. In several BHXBs, these have been in alignment. For GS~1354$-$64, past radio detections were unresolved and did not yield a robust jet position angle (e.g., \citealt{2001MNRAS.323..517B}). 

During its most recent 2025--2026 outburst, GS~1354$-$64 underwent a rapid increase in X-ray flux, with the 2--10~keV luminosity rising sharply over a period of days \citep{2026ATel17618....1N}. The Monitor of All-sky X-ray Image (\maxi) Gas Slit Camera (GSC) 
%aboard the International Space Station 
\citep{2009PASJ...61..999M} revealed this flare reaching $\gtrsim500$~mCrab with a slight spectral softening. The system subsequently declined over the next $\sim$1--2-days. This behavior is consistent with a brief excursion to the soft-intermediate or steep-power-law state~\citep{2006ARA&A..44...49R}. 

A radio monitoring campaign was initiated with the MeerKAT radio telescope \citep{2016mks..confE...1J} at 1.28 GHz as part of the X-KAT program (PI Fender; SCI-20230907-RF-01). Observations were performed on 2026 January 4, January 26, and February 9, yielding flux densities of $3.66\pm0.03$~mJy \citep{2026ATel17582....1S}, $13\pm1.3$~mJy \citep{2026ATel17638....1S}, and $63\pm3$~mJy (Salvaggio et al. in prep) respectively, indicating a bright radio flare.

In this Letter, we present the first X-ray polarimetric observation of GS~1354$-$64, obtained by \ixpe between episodes of flared X-ray activity during its ongoing outburst. We present model-independent polarimetric analysis, spectropolarimetric analysis jointly with a simultaneous \nustar observation of the source, and show a quasi-periodic oscillation (QPO) seen in timing analysis of the \ixpe data. We discuss the implications of these analyses on the geometry and evolution of the source during its outburst.

\section{Observations and Data Reduction}

\subsection{Observations}
\subsubsection{\maxi Monitoring}

We monitored GS~1354$-$64 using \maxi and obtain per-orbit light curves in the 2--4, 4--10, and 10--20~keV energy bands covering the interval 2025 December 29--2026 February 19 (MJD~61038--61090), encompassing the source's early evolution during the 2025--2026 outburst.

\subsubsection{\ixpe}

\ixpe observed GS~1354$-$64 during the period 2026 February 07--09 (ObsID 05002101), with total effective exposure of $\sim$150~ks per Detector Unit (DU). Due to the ongoing \ixpe detector anomaly calibration, we exclude DU2 data from our analysis.

We select a circular source region of radius 60\arcsec\ from the \ixpe Level 2 event files using \texttt{SAOImage DS9} \citep{2003ASPC..295..489J}, centered via ``centroid'' feature. Per the prescription outlined in \cite{2023AJ....165..143D}, we apply no background rejection or subtraction given the high source count rate per DU of 7.30~$\mathrm{s^{-1}arcmin^{-2}}$. We use time-dependent response version 20250701, and bin the spectra into 200~eV intervals (oversampling the resolution by a factor $\sim$2--3).

\subsubsection{\nustar}

GS~1354$-$64 was observed with the Nuclear Spectroscopic Telescope Array (\nustar), with simultaneous overlap of the \ixpe observation, on 2026~February~07 (ObsID 81202301002), as marked by the gray shading of Fig.~\ref{fig:1354_lc}. The observation has a total effective exposure of $\sim$15.6~ks per focal plane module (FPM). We generate cleaned event files with \texttt{nupipeline} (using the latest CALDB dated 20260210) and apply standard screening criteria (e.g., saamode=optimized, tentacle=yes). We extract circular source regions of radius 120\arcsec\ using \texttt{SAOImageDS9} \citep{2003ASPC..295..489J}, centered as described previously. Background spectra of equivalent radius are obtained from source-free regions per detector. Response files are produced using \texttt{nuproducts}. The FPMA and FPMB spectra are optimally grouped \citep{2016A&A...587A.151K}, additionally imposing a minimum of 30 counts per bin, and fitted over the 3--79~keV range.

\subsection{Data Reduction}

\subsubsection{Model-Independent Polarimetric Analyses}

We perform model-independent polarimetric analysis using both the \texttt{ixpeobssim} (version 31.0.3) weighted polarization cube (PCUBE) method following the SIMPLE weighting prescription\footnote{Weighted and unweighted PCUBE results agree within $1\sigma$.} \citep{2015APh....68...45K, 2022SoftX..1901194B}, and the event-based Bayesian framework \texttt{QUEEN-BEE} (version 3.2) using its constant polarization model \texttt{scout\_const} \citep{ravi2025queenbee, 2026ApJ...997...60R}. 

We additionally perform both binned PCUBE and event-based QUEEN-BEE analysis of smooth polarization variability, probing both continuous time- and energy-dependent PA rotation across the observation. With QUEEN-BEE, we compare both the time-rotating model \texttt{scout\_rot} and the energy-rotating model \texttt{scout\_energy\_rot} against a constant PA model and the unpolarized null hypothesis. 

\subsection{\ixpe Timing Analysis}

We perform timing analysis of the \ixpe data using the Python timing package Stingray \citep{2019ApJ...881...39H, matteo_bachetti_2023_7970570} version 1.1.2.4. Events from DU1 and DU3 are filtered into good-time intervals, and high-time-resolution light curves generated with a time bin of $\Delta t =1~\mathrm{ms}$ yielding a Nyquist frequency of 500~Hz. A power density spectrum (PDS) is computed averaging over 100~s segments and shown in fractional-rms normalization. Deadtime effects were corrected using the Fourier Amplitude Difference (FAD) method \citep{2018ApJ...853L..21B}. The resulting power spectrum was logarithmically rebinned with 5\% fractional width. The Poisson-noise level was estimated from the mean power above 100 Hz.

\section{Results}

\subsection{The IXPE Observation}

\begin{figure}
    \centering
    \includegraphics[width=0.45\textwidth, trim=0 0 0 5, clip]{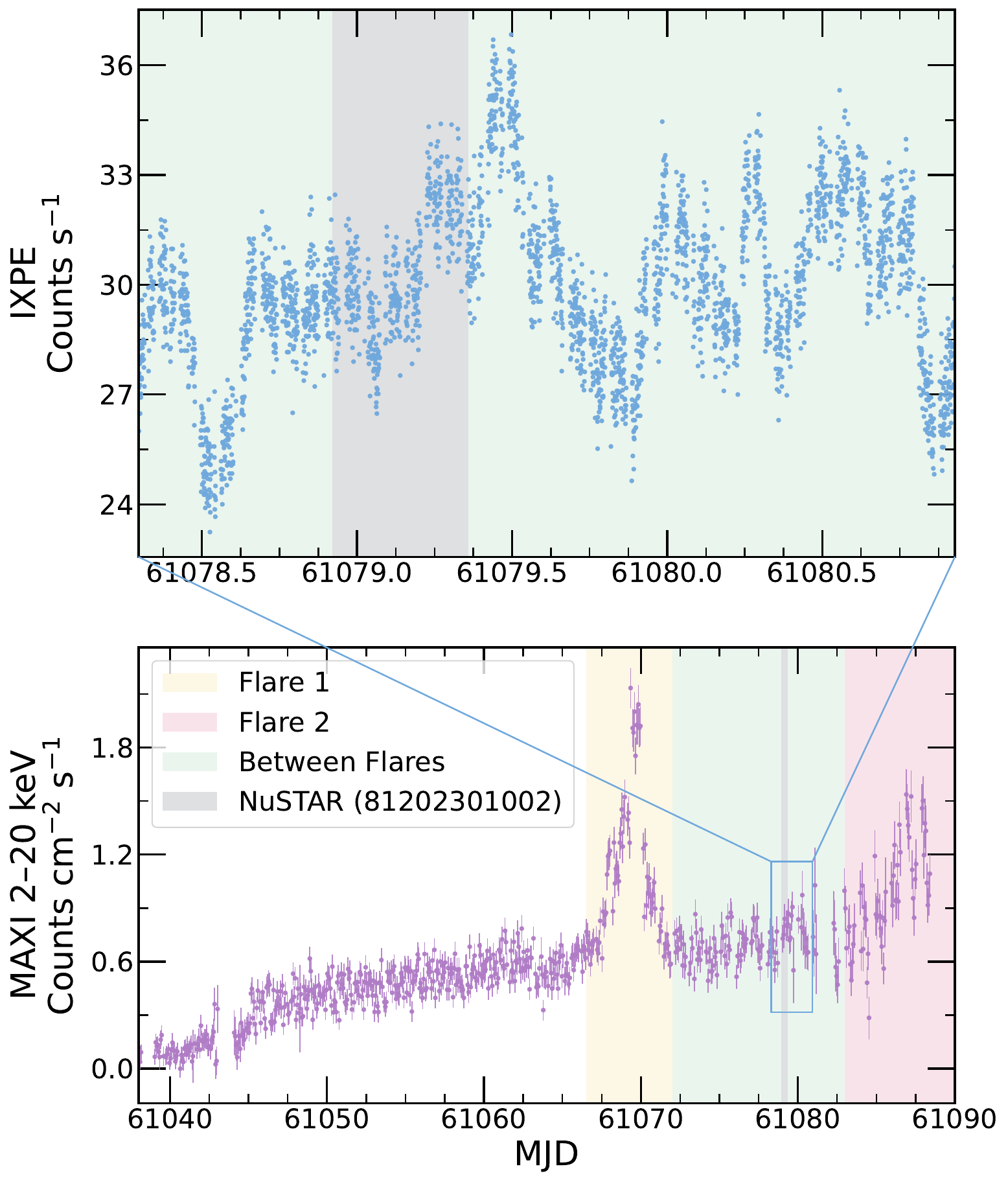}
    \vspace{-0.3em}
    \caption{\maxi/GSC light curve of GS~1354$-$64 during its 2025-2026 outburst (MJD~61038--61090). The boxed region in the 2--20~keV light curve depicts the \ixpe observation and the gray shading denotes the simultaneous \nustar observation. The \ixpe 2--8~keV light curve is shown in detail in the top panel, from the combined DU1+DU3 event list and binned at 50 s.}
    \label{fig:1354_lc}
\end{figure}

As shown in Fig.~\ref{fig:1354_lc}, the \ixpe observation occurred shortly after a pronounced $\gtrsim500$~mCrab X-ray flare had declined back to $\sim200$mCrab. The \ixpe light curve (Fig.~\ref{fig:1354_lc}) exhibits strong red-noise variability throughout the observation, while the hardness ratio (computed from 4--8~keV~$/$~2--4~keV) remains stable within $\sim$10$\%$ over the observation.

\begin{figure}
    \centering
    \includegraphics[width=0.45\textwidth, trim=0 5 0 5, clip]{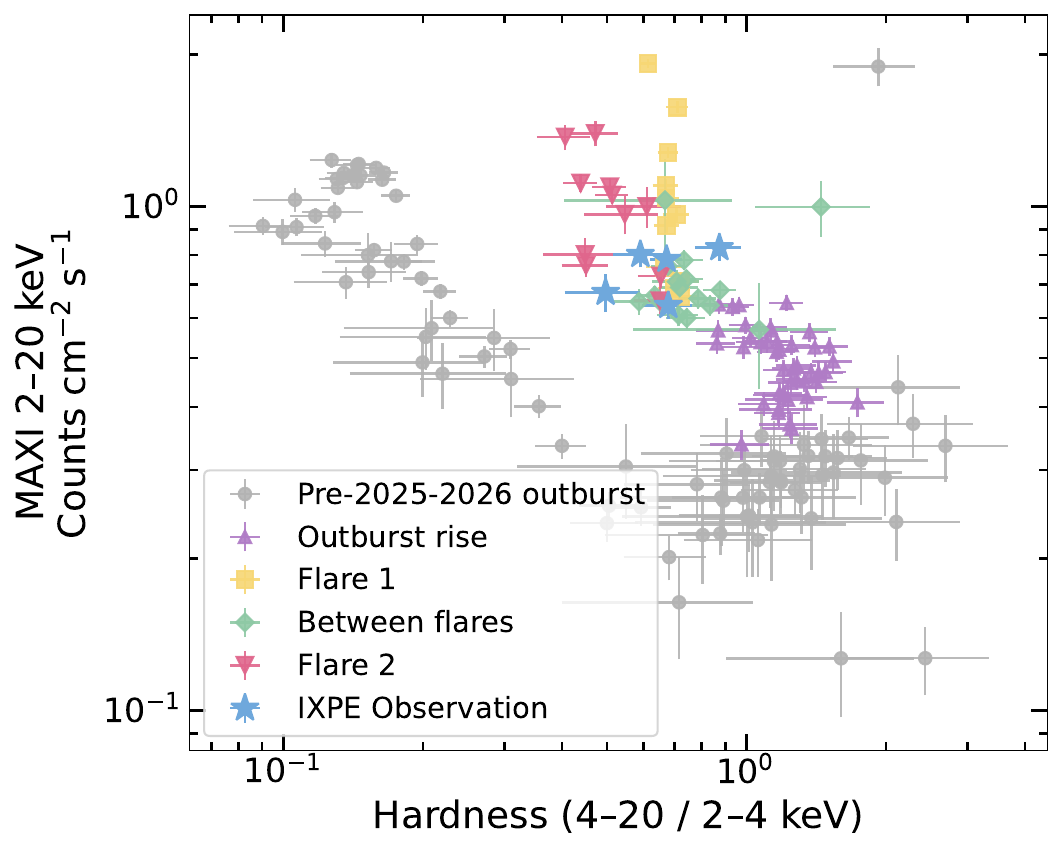}
    \vspace*{1em}
    \caption{\maxi/GSC HID of GS~1354$-$64 from MJD 55053 to 61090, using 12-hour intervals. Colored points illustrate the ongoing 2025--2026 outburst, with colors corresponding to outburst phase; the \ixpe observation interval is marked with star symbols.}
    \label{fig:1354_hid}
\end{figure}

    The hardness-intensity diagram (HID; Fig.~\ref{fig:1354_hid}), the level of variability, the presence of a type-C QPO, and the source's spectral properties establish the source in an intermediate state at the time of the observation (see Sects.~\ref{subsec:timing}--\ref{subsec:specpol}).

\subsection{Polarimetry}

\begin{figure*}
    \centering
    
    \includegraphics[width=0.49\textwidth, trim=0 45 0 5, clip]{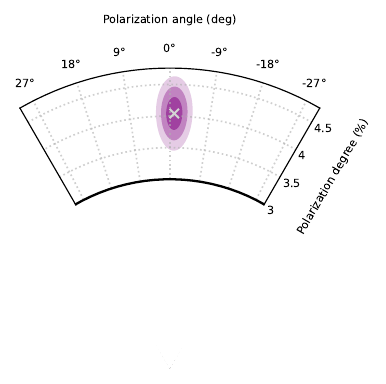}
    \hfill
    \includegraphics[width=0.49\textwidth]{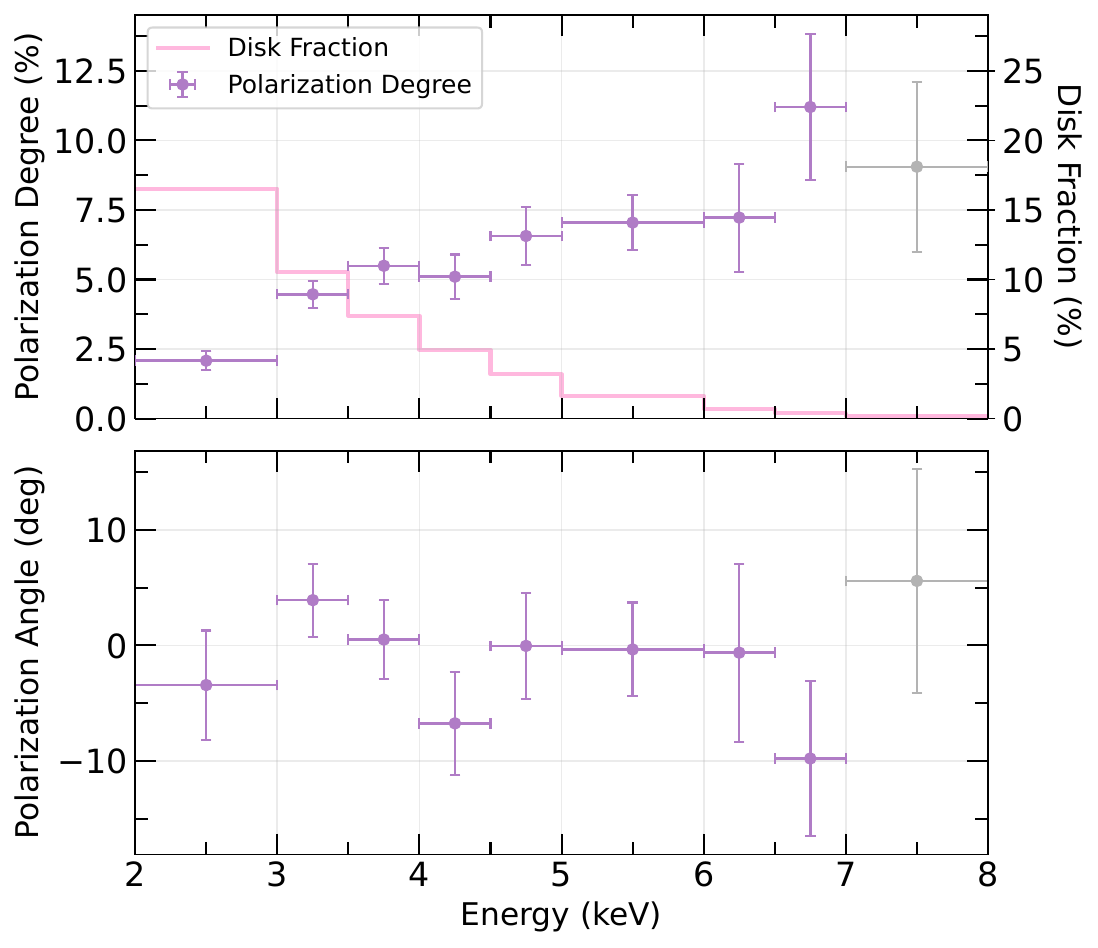}

    \caption{
    Polarimetric summary of the GS~1354$-$64 \ixpe\ observation.
    \textbf{Left:} Constant-PA (\texttt{scout\_const}) QUEEN-BEE polar plot. Contours represent $68\%, 95\%,~\mathrm{and}~99.7\%$ credible intervals.  
    \textbf{Right:} Energy-dependent PD and PA behavior obtained from binned PCUBE analysis. Error bars denote $1\sigma$ uncertainties. Points in grey denote bins with $\textless3\sigma$ significance. The corresponding disk fraction---defined as the ratio of disk flux to total (disk + Comptonized) flux---is overplotted in pink.}
    \label{fig:1354_polar}
\end{figure*}

We measure significant polarization of the source from 2--8~keV under both model-independent polarimetric approaches. PCUBE yields a PD of $4.0\pm0.3\%$ and PA of $-1\degr\pm2\degr$ ($1\sigma$ uncertainties), at a significance of $14.5\sigma$. QUEEN-BEE \texttt{scout\_const} returns consistent polarization estimates to PCUBE (Fig.~\ref{fig:1354_polar}), yielding a PD of $4.0\pm0.2\%$ and a PA of $-1.0^{+1.7}_{-1.6}\degr$ ($90\%$ credible intervals), with a log Bayes Factor $\Delta \mathrm{ln}(Z)=283\pm1$ between the constant polarization model and the null unpolarized hypothesis. 

Energy-binned PCUBE analysis (right panel of Fig.~\ref{fig:1354_polar}) indicates a significant increase in PD as a function of energy. One expects any fluorescent Fe-K emission to be depolarizing, but here, with a Fe-K flux contribution of $< 10\%$ in the 6--7~keV bin, we are not sufficiently sensitive to detect this directly. The PA appears constant across the energy band. Consistent with this, QUEEN-BEE analysis using \texttt{scout\_energy\_rot} identifies no evidence for PA rotation with energy. Model comparison yields $\Delta \mathrm{ln}(Z)=-5\pm1$ favoring a constant PA model, with the best-fit rotation rate of 
$0.0^{+1.7}_{-1.5}$~$\mathrm{degrees}/\mathrm{keV}$ ($90\%$ credible interval). Further, no significant rotation of the PA with time was found using the QUEEN-BEE function \texttt{scout\_rot}: $\Delta \mathrm{ln}(Z)=-3.0\pm1.2$ between the time-rotating PA model and the constant PA model, offering moderate evidence in favor of the constant PA model \citep{Jeffreys1986}. 

\subsection{Timing}
\label{subsec:timing}
\begin{figure}
    \centering
    \includegraphics[width=0.45\textwidth, trim=0 5 0 5, clip]{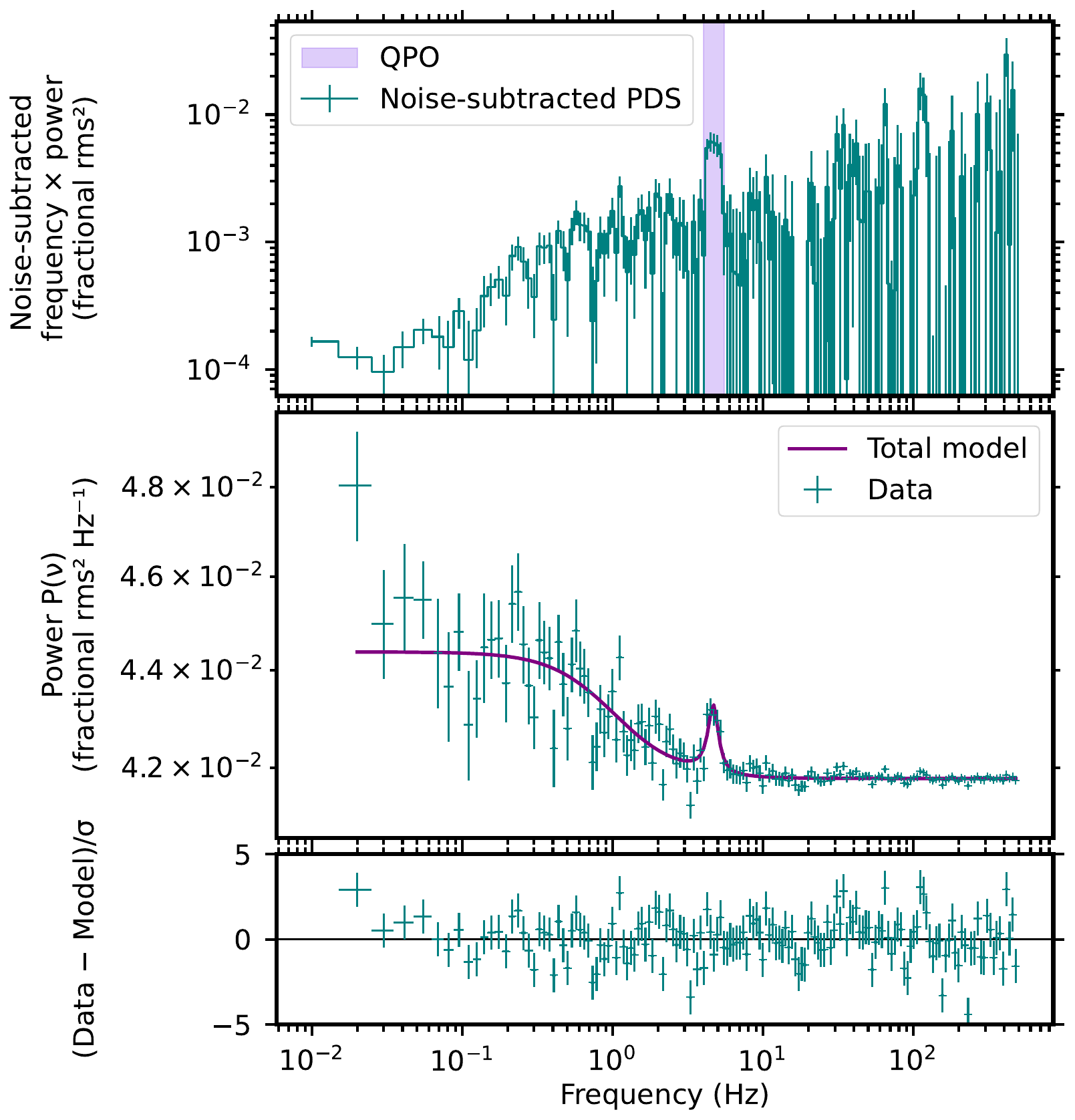}
    \vspace{-0.7em}
    \caption{Timing properties of GS~1354$-$64 from \ixpe. \textbf{Top:} Noise-subtracted PDS plotted as $\nu(P(\nu)-P_{\rm noise})$ (fractional rms$^2$), with the shaded region indicating the QPO band. \textbf{Middle:} The rms-normalized power spectrum $P(\nu)$ (fractional rms$^2$ Hz$^{-1}$) with the best-fit model of a constant and two Lorentzians overplotted. \textbf{Bottom:} Fit residuals.
    }\label{fig:1354_timing}
\end{figure}

The deadtime-corrected, logarithmically-rebinned PDS (Fig.~\ref{fig:1354_timing}) shows strong broadband variability with a flat-top noise component that steepens above $\sim$1--2~Hz. Superimposed on this continuum is a statistically significant QPO, highlighted in the figure. The PDS is well-described by a model consisting of a constant, representing the Poisson noise level, plus two Lorentzian components (see Table \ref{tab:gs1354_timing}). One Lorentzian describes the QPO, with the best fit yielding a QPO centroid frequency of $\nu_o=4.7\pm0.1$~Hz and width of $0.9^{+0.3}_{-0.2}$~Hz, corresponding to a quality factor $Q\approx5.2$. The integrated fractional rms amplitude of the QPO is $4.4\% \pm 0.5\%$. The second Lorentzian is zero-centered and describes the broadband noise, with a width of $1.9^{+0.4}_{-0.3}$~Hz. The fit statistic is $\chi^2=255.4$ for 153 degrees of freedom, and $\chi^2$/d.o.f.$=356.22/156$ when fitting without the QPO component. The QPO is clearly detected above the local continuum, with excess power confined to the $\sim$4--6~Hz band (highlighted in Fig.~\ref{fig:1354_timing}). A candidate feature appears near $\sim$100~Hz, which we believe is likely an artifact of the instrumental deadtime correction (see Fig.~2 of \citealt{2018ApJ...853L..21B}). The QPO centroid frequency and broadband noise morphology are characteristic of Type-C QPOs observed in numerous BHXBs (e.g., \citealt{2005ApJ...629..403C, 2010LNP...794...53B, 2019NewAR..8501524I}).

\begin{deluxetable}{lccc}
\tablecaption{Best-fit PDS parameters for GS~1354$-$64 using the model: 
$\mathrm{powerlaw} + \mathrm{lorentz} + \mathrm{lorentz}$ (the powerlaw index is frozen at 0 to act as a constant, to describe the Poisson noise). 
Errors are reported at the 90\% confidence level ($\Delta\chi^2=2.706$ for one parameter).
\label{tab:gs1354_timing}}
\tablehead{
\colhead{Component} &
\colhead{Parameter} &
\colhead{Value} &
\colhead{Unit}
}
\startdata
powerlaw & Photon Index & 0 (frozen) & - \\
powerlaw & Norm & $0.04179 \pm 0.00001$ & - \\
\hline
\multicolumn{4}{c}{\textit{Lorentzian 1 (QPO)}} \\
\hline
lorentz & Centroid ($\nu_0$) & $4.7\pm0.1$ & Hz \\
lorentz & Width & $0.9_{-0.2}^{+0.3}$ & Hz \\
lorentz & Norm & $(1.9\pm0.4) \times 10^{-3}$ & - \\
\hline
\multicolumn{4}{c}{\textit{Lorentzian 2 (Broadband Noise)}} \\
\hline
lorentz & Centroid ($\nu_0$) & 0 (frozen) & Hz \\
lorentz & Width & $1.9_{-0.3}^{+0.4}$ & Hz \\
lorentz & Norm & $(3.9\pm0.5) \times 10^{-3}$ & - \\
\hline
\multicolumn{4}{l}{$\chi^2/\mathrm{d.o.f.} = 255.36/153$}
\enddata
\end{deluxetable}

\begin{deluxetable*}{lccc}
\tablecaption{Best-fit spectral parameters for GS~1354$-$64 from the joint \ixpe (Stokes I) and \nustar fit using 
$\mathrm{TBabs} \times [\mathrm{diskbb} + (\mathrm{nthComp} \times\mathrm{smedge}) + \mathrm{laor}]$. With \textsc{laor}, for simplicity, we adopt the default Thorne-limit Kerr disk ($a_\ast=0.998$). In the fit we fix the emissivity index to $q=3$ and the line energy to 6.7~keV (Fe~\textsc{xxv}); only its normalization and inclination are free. For \textsc{smedge}, we fix the width at 7 keV and index at its default value $–2.67$. Polarimetric parameters obtained from both the ``orthogonal components'' and ``Compton-increasing'' scenarios are included beneath.  
Errors are reported at the 90\% confidence level.
\label{tab:gs1354_spec}}
\tablehead{
\colhead{Component} &
\colhead{Parameter} &
\colhead{Value} &
\colhead{Unit}
}
\startdata
\multicolumn{4}{c}{\textbf{Spectral Fit (\ixpe I$+$\nustar)}}\\
\hline
TBabs & $N_{\rm H}$ & 0.7 (frozen) & $10^{22}\,\mathrm{cm^{-2}}$ \\
diskbb & $T_{\rm in}$ & $0.61^{+0.01}_{-0.02}$ & keV \\
diskbb & Norm & $860\pm60$ & - \\
nthComp & $\Gamma$ & $2.462^{+0.002}_{-0.005}$ & - \\
nthComp & $kT_e$ & $50^{+20}_{-10}$ & keV \\
nthComp & Norm & $2.01^{+0.05}_{-0.09}$ & - \\
smedge & $E_{\rm edge}$ & $7.00^{+0.01}_{-7.00}$ & keV \\
smedge & $\tau_{\rm max}$ & $0.72^{+0.01}_{-0.03}$ & - \\
laor & $q$ & 3 (frozen) & - \\
laor & $E_{\rm line}$ & 6.7 (frozen) & keV \\
laor & Inclination & $80 \pm 6$ & deg \\
laor & Norm & $(3^{+4}_{-2})\times10^{-4}$ & - \\
\hline
Gain & Slope & $0.982^{+0.003}_{-0.001}$ & - \\
Gain & Offset & $0.072^{+0.007}_{-0.005}$ & keV \\
\hline
$\chi^2/\mathrm{d.o.f.} = 720.72/489$ \\
\hline
\multicolumn{4}{c}{\textbf{Polarimetric Fit (\ixpe Q$+$U; ``Orthogonal Components'' case)}} \\
\hline
polconst (disk) & $A_{\rm disk}$ & $0.27 \pm 0.07$ & - \\
polconst (disk) & $\psi_{\rm disk}$ & $\psi_{C}+90^\circ$ (tied) & deg \\
polconst (nthComp) & $A_{C}$ & $0.079 \pm 0.009$ & - \\
polconst (nthComp) & $\psi_{C}$ & $-1.1 \pm 2.5$ & deg \\
\hline
$\chi^2/\mathrm{d.o.f.} = 116.85/117$ \\
\hline
\multicolumn{4}{c}{\textbf{Polarimetric Fit (\ixpe Q$+$U; ``Compton-Increasing’’ case with PD$_{\rm {\bf disk}}$=0\%)}} \\
\hline
polconst (disk) & $A_{\rm disk}$ & 0 (frozen) & - \\
pollin (nthComp) & $A_{1}$ & $0.009^{+0.011}_{-0.009}$ & - \\
pollin (nthComp) & $A_{\rm slope}$ & $0.015 \pm 0.004$ & - \\
pollin (nthComp) & $\psi_{C}$ & $-1.1 \pm 2.5$ & deg \\
\hline
$\chi^2/\mathrm{d.o.f.} = 120.34/117$ \\
\hline
\multicolumn{4}{c}{\textbf{Polarimetric Fit (\ixpe Q$+$U; ``Orthogonal Components'' case with $\boldsymbol{\mathrm{PD_{\rm {\bf disk}}}=6\%}$)}} \\
\hline
polconst (disk) & $A_{\rm disk}$ & 0.06 (frozen) & - \\
polconst (nthComp) & $A_{C}$ & $0.054 \pm 0.004$ & - \\
polconst (nthComp) & $\psi_{C}$ & $-1.2 \pm 2.6$ & deg \\
\hline
$\chi^2/\mathrm{d.o.f.} = 141.76/118$ \\
\hline
\multicolumn{4}{c}{\textbf{Polarimetric Fit (\ixpe Q$+$U; ``Compton-Increasing’’ case with $\boldsymbol{\mathrm{PD_{\rm {\bf disk}}}=6\%}$)}} \\
\hline
polconst (disk) & $A_{\rm disk}$ & 0.06 (frozen) & - \\
pollin (nthComp) & $A_{1}$ & $0.024\pm0.011$ & - \\
pollin (nthComp) & $A_{\rm slope}$ & $0.012\pm0.004$ & - \\
pollin (nthComp) & $\psi_{C}$ & $-1.1 \pm 2.5$ & deg \\
\hline
$\chi^2/\mathrm{d.o.f.} = 118.77/117$ \\
\hline
\enddata
\end{deluxetable*}

\subsection{Spectropolarimetry}
\label{subsec:specpol}
\begin{figure*}[t]
    % \vspace*{0.8em}
    \centering
    
    \includegraphics[width=0.49\textwidth, trim=0 5 0 5, clip]{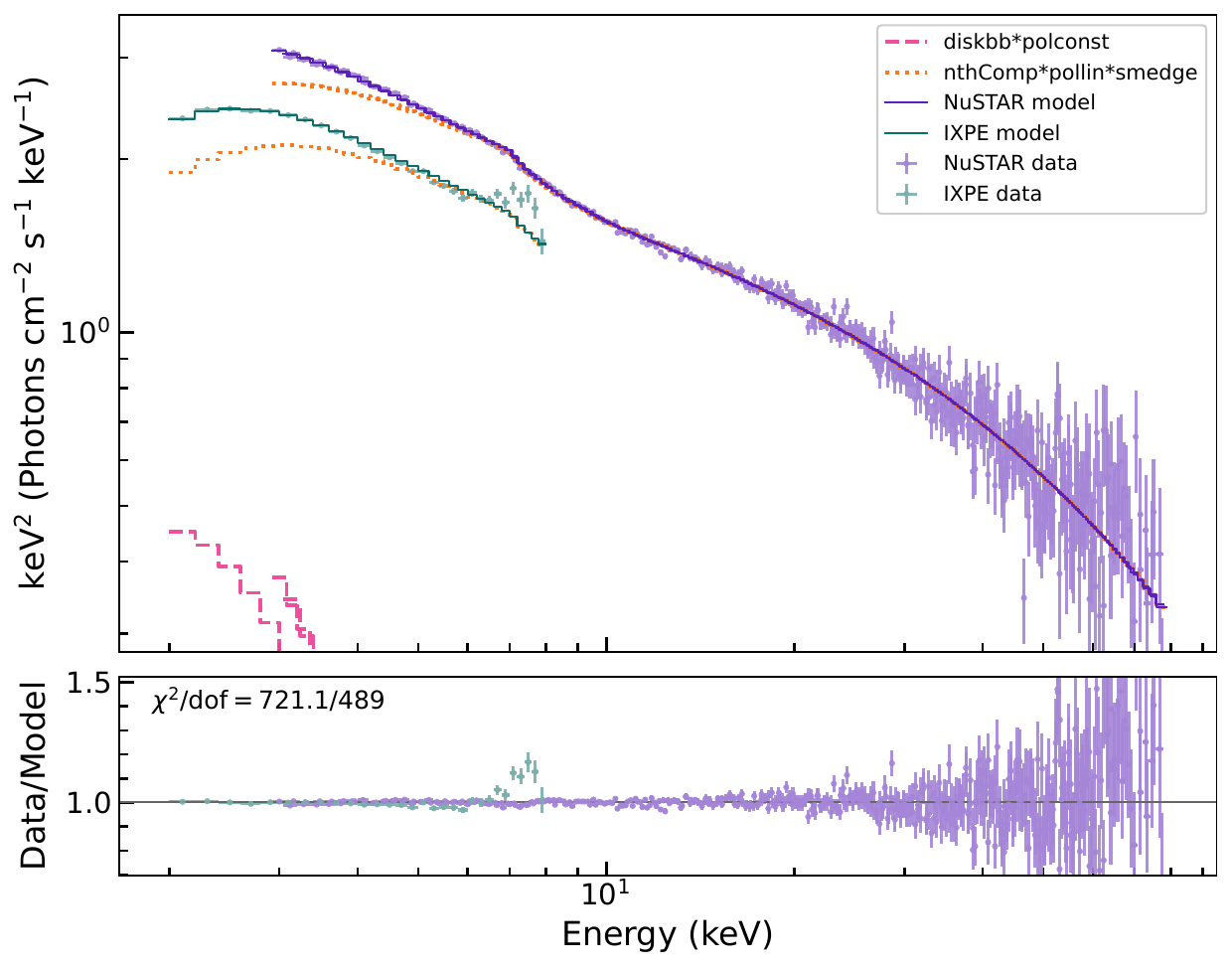}
    \includegraphics[width=0.49\textwidth, trim=0 5 0 5, clip]{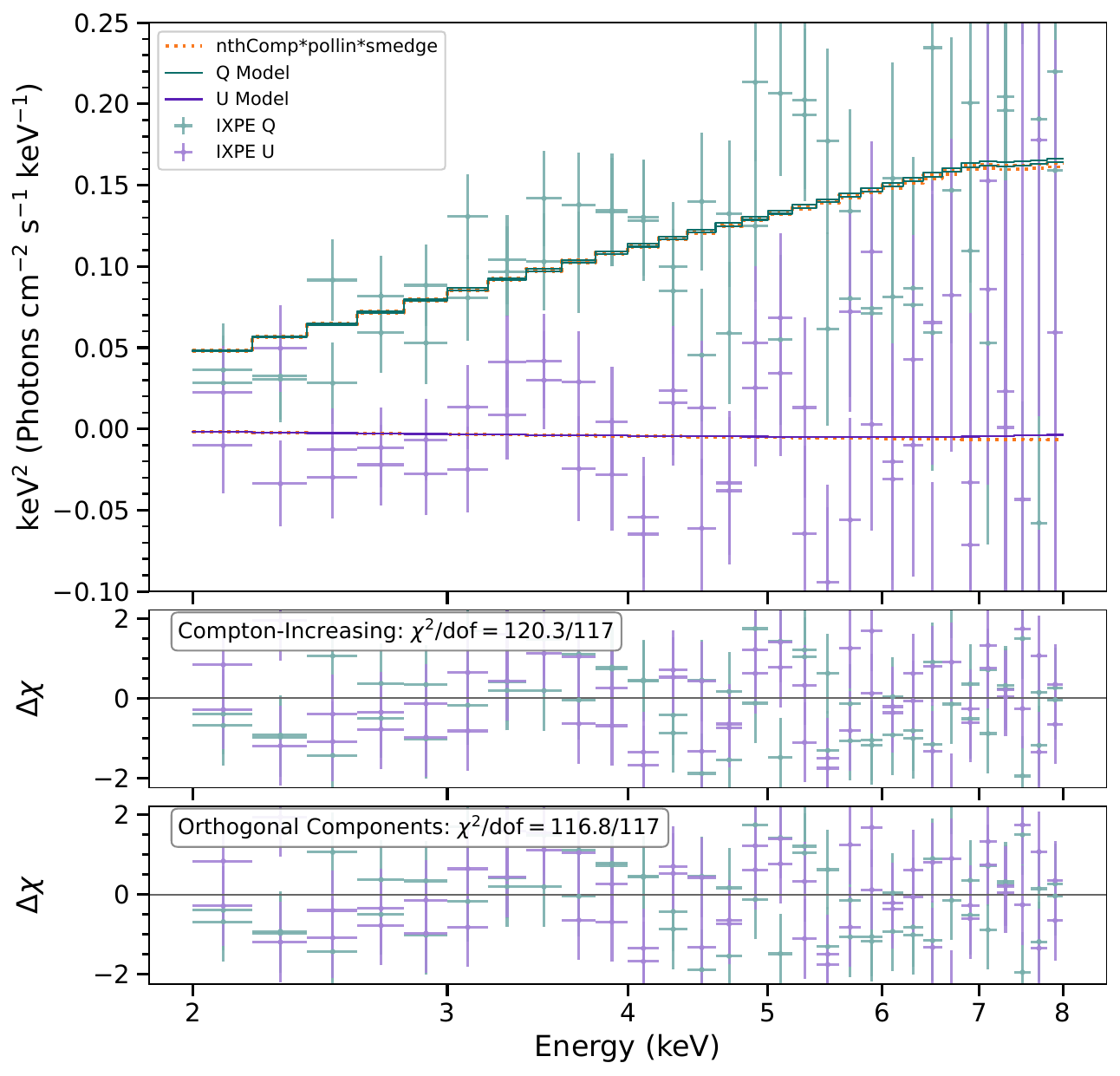}
    
    \caption{
    Joint \ixpe + \nustar spectral fitting for GS~1354$-$64. 
    \textbf{Left:} \ixpe I + \nustar spectra and best-fitting model with fit residuals. The model components are shown separately, but the minimal laor contribution falls below the scale of the plot. 
    \textbf{Right:} \ixpe Q + \ixpe U spectra and best-fitting ``Compton-increasing'' model with fit residuals. The baseline ``Compton-increasing'' model assumes an unpolarized disk and a Comptonizing component linearly trending PD. The fit residuals for the formally preferred ``orthogonal components'' model are included in the bottommost panel.}
    \label{fig:1354_spectra}
\end{figure*}

To jointly model the spectral and polarization properties, we first performed a broadband spectral fit to the \ixpe Stokes I spectra (DU1 and DU3) together (in the 2--8~keV range) with the contemporaneous \nustar FPMA/FPMB spectra (in the 3--79~keV range) using \texttt{XSPEC} version 12.14.0h \citep{1996ASPC..101...17A}. We adopt the model \textsc{TBabs$\times$(diskbb$+$nthComp)}. We add additional phenomenological components to capture residual reflection-related edge and line features by slightly modifying the Compton component: \textsc{smedge} \citep{1991PhDT........55E} and \textsc{laor} \citep{1991ApJ...376...90L} (see Fig.~\ref{fig:1354_spectra} and Table~\ref{tab:gs1354_spec}). For \textsc{laor}, we adopted the default maximal-spin Kerr geometry ($a_\ast=0.998$; Thorne limit) and fixed the emissivity index to $q=3$ and the line energy to 6.7~keV (Fe~\textsc{xxv}). For \textsc{smedge}, the edge width (7 keV) and index ($–2.67$) were fixed. Interstellar absorption was modeled using TBabs, adopting the abundances of \cite{2000ApJ...542..914W} and photoelectric cross sections of \cite{1996ApJ...465..487V}. We additionally modeled detector gain offsets for \ixpe DU1 and DU3 using the XSPEC \texttt{gain fit} command, allowing both slope and offset to vary during the Stokes-I fit (but tying parameters for DU1 and DU3 together after they were found to be statistically compatible). The best-fit gain slope was $0.982^{+0.003}_{-0.001}$, with offsets of $72^{+7}_{-5}$~eV. We fixed $N_{\rm H}=7\times10^{21}{\rm\,cm^{-2}}$ (see \citealt{2016MNRAS.460..942K}) %and the electron temperature $kT_e=50{\rm\,keV}$
to stabilize the continuum decomposition. The parameters of the reflection components (\textsc{smedge} and \textsc{laor}) were held fixed, while only their normalization (and inclination, in the case of \textsc{laor}) were allowed to vary. In \textsc{nthComp}, we adopted a disk seed spectrum linked to the \textsc{diskbb} component's temperature. 
After obtaining an acceptable joint \nustar$+$\ixpe Stokes-I fit, we froze all spectral parameters and \ixpe's gain parameters and then fit the \ixpe Stokes Q and U spectra to constrain the polarization. For the spectropolarimetric analysis, we explored two physical scenarios using the same frozen spectral decomposition:

\begin{enumerate}
    \item An ``orthogonal components'' case in which the disk and Comptonized emission each has a constant (with energy) PD but with components constrained to have PAs separated by $90^\circ$.
    \item A ``Compton-increasing'' case in which the disk is assumed unpolarized and the Compton component has a PD allowed to vary linearly with energy across the \ixpe bandpass.% (e.g. with \textsc{pollin}.)
\end{enumerate}

In the ``orthogonal components'' picture, the data prefer an extremely polarized disk with ${\rm PD}_{\rm disk}=0.27\pm0.07$ and Compton component (${\rm PD}_{\rm C}=0.079\pm0.009$), yielding a fit with $\chi^2/\mathrm{d.o.f.}=116.85/117$ (see Table~\ref{tab:gs1354_spec} and right panel of Fig.~\ref{fig:1354_spectra}).  Such a high PD isn't expected for standard accretion disks (e.g., see \citealt{1960ratr.book.....C}) and so we view this fit as implausible. To investigate the possibility of a more reasonable disk polarization, we explore freezing ${\rm PD}_{\rm disk}=0.06$ which yields a significantly worse fit $\Delta \chi^2=24.9$ for one fewer degree of freedom (see Table~\ref{tab:gs1354_spec}). The ``Compton-increasing'' picture with an unpolarized disk and a Comptonized component with linearly increasing polarization provides a slightly worse but comparable fit, with $A_{1}=0.009^{+0.011}_{-0.009}$, $A_{\rm slope}=0.015\pm0.004$, and $\chi^2/\mathrm{d.o.f.}=120.34/117$. Here $A_1$ is the PD at 1 keV and $A_{\rm slope}$ is the change in PD per keV. We again freeze ${\rm PD}_{\rm disk}=0.06$, which yields a marginally better fit $\Delta \chi^2=1.57$ (Table~\ref{tab:gs1354_spec}). The coronal polarization has a modestly shallower increase of $A_{\rm slope}=0.012\pm0.004$ and a somewhat higher reference PD at 1~keV of $0.024\pm0.011$. In every case, the inferred polarization angle of the dominant coronal component is consistent with PA=$\psi_{1}\simeq-1^\circ$. 

\section{Discussion}

We present the first X-ray polarimetric observation of GS~1354$-$64, performed by \ixpe, capturing the source in an intermediate state between days-long flares in the X-ray intensity while near the cusp of state transition. The observation provides both an opportune polarimetric snapshot of a system having stalled after a failed state transition and offers an addition to the sparse sample of intermediate-state BHXBs observations with \ixpe (e.g., Swift~J1727.8$-$1613, \citealt{2023ApJ...958L..16V,2024A&A...686L..12P}; GX 339--4, \citealt{2025ApJ...978L..19M}; and IGR~J17091$-$3624, \citealt{2025MNRAS.541.1774E}). This offers a valuable opportunity to probe how accretion geometry reorganizes---or fails to reorganize---during this critical phase.

We measure $\sim$4\% polarization across the 2--8~keV band, with PD increasing near-monotonically across the bandpass from $2.1\pm0.3\%$ in the 2--3~keV band to $11\pm3\%$ in the 6.5--7~keV band. We find no significant trend of the PA with energy. Energy-dependent PA rotation is generally expected if distinct spectral components (e.g., disk and coronal emission) contribute with different intrinsic PD and PA or if relativistic returning radiation induces strong energy-dependent rotations \citep{2009ApJ...701.1175S, 2010ApJ...712..908S}. The lack of monotonic PA rotation with energy generally favors a single geometric configuration dominating the polarized emission across the bandpass, consistent with a geometrically stable inner flow over the duration of the observation. Non-linear or weak trends of PA with energy may still be present but are undetected. The consistency in PA across the \ixpe bandpass matches observations of other systems observed including across state transition (e.g., Swift~J1727.8$-$1613) where PD is seen to evolve, but PA remains stable across states \citep{2024ApJ...966L..35S, 2024ApJ...968...76I}.

GS~1354$-$64 joins the growing collection of BHXBs exhibiting an apparent trend of PD increasing with energy, as seen, e.g., in Cyg~X-1 \citep{2022Sci...378..650K, 2024ApJ...969L..30S}, IGR~J17091$-$3624 \citep{2025MNRAS.541.1774E}, Swift~J1727.8$-$1613 \citep{2024ApJ...968...76I}, 4U~1630$-$47 \citep{2024ApJ...964...77R}, and LMC~X-3 \citep{2024ApJ...960....3S}. Among these sources, GS~1354$-$64 exhibits the largest amplitude increase of PD of $\sim$9\% across 2--8~keV alongside IGR~J17091$-$3624 ($\sim$8\%, but not a significant trend, as reported in \citealt{2025MNRAS.541.1774E}) compared to $\sim$2.5\% \citep{2022Sci...378..650K} and $\sim$2\% \citep{2024ApJ...969L..30S} for Cyg~X-1, $\sim$5\% for 4U~1630$-$47 \citep{2024ApJ...964...77R}, and $\sim$5\% for LMC~X-3 in \citep{2024ApJ...960....3S}. Moreover, the energy-resolved PD analysis of GS~1354$-$64 appears consistent with a linear increase across the bandpass unlike the near-flat 2--7~keV PD ($\sim$5\%) and steep-jump in 7--8~keV PD ($\sim$12\%) observed in the dim hard state of Swift~J1727.8$-$1613 \citep{2024A&A...686L..12P}.

Our joint spectropolarimetric modeling explores two major scenarios to explain this PD trend with energy. In the ``orthogonal components'' picture, the disk and Comptonized emission are assumed to each have a constant polarization across the \ixpe bandpass, but with PA rotated $90^\circ$ relative to each other. In the ``Compton-increasing'' case, we assume the disk is unpolarized while the Comptonized component exclusively carries the polarization which can grow (or decline) with energy.

The ``orthogonal components'' picture is formally preferred with $\Delta\chi^2=3.5$ for the same degrees of freedom.  However, the best fit implies a physically implausible disk PD exceeding 20\%. This is to be compared with a $\sim11\%$ upper bound for an electron scattering atmosphere observed completely edge on \citep{1960ratr.book.....C}. As eclipsing and other obscuration would occur at extreme inclinations, we adopt a more reasonable approximation for a maximum polarization for a highly-inclined disk of $\sim 6\%$. Imposing this, we find the fit is substantially worse ($\Delta\chi^2=24.9$ for one fewer degree of freedom). We find that a much more physically plausible interpretation arises from the slightly-worse fit (statistically) provided by the ``Compton-increasing'' picture, in which an unpolarized disk and a rapidly growing PD (with energy) from the corona successfully describes the Stokes Q and U data.

In this scenario, the best-fit parameters suggest the Comptonized component maintains a $\sim$1\% PD at 1~keV with a $\sim$1.5~\%/keV growth in PD with energy. Here, higher-energy photons have likely undergone more scatterings, emerge with a more anisotropic angular distribution and therefore a higher PD (e.g., \citealt{1985A&A...143..374S}; \citealt{1996ApJ...470..249P}; see also scattering-order decompositions in \citealt{2018A&A...619A.105T}). The increasing PD therefore reflects the growing dominance and/or increasing scattering order within a geometrically coherent hot inner flow, consistent with a radially-extended Comptonizing region characteristic of an intermediate state. However, neither of the two scenarios, which can each account for the PD increase with energy in GS~1354$-$64, offers a natural explanation for the ubiquity of the signature increase in PD with energy in BHXRBs even in sources, e.g., LMC~X-3 \citep{2024ApJ...960....3S} and 4U~1630$-$47 \citep{2024ApJ...964...77R}, that lack significant Comptonized emission. 

Accretion disk winds, when present, may complicate the interpretation of disk and coronal geometries inferred from spectropolarimetry. Radiatively driven or magnetically launched winds can scatter and reprocess X-ray photons altering both the PD and PA and potentially mimicking or modifying signatures otherwise attributed to the inner disk or corona \citep{2024MNRAS.527.7047T, 2025A&A...694A.230N}. In particular, scattering in an equatorial or clumpy wind may rotate the PA or enhance/suppress the net PD depending on viewing geometry and optical depth. While GS~1354$-$64 does not show evidence of wind features in our fits, this can be more precisely ascertained with high-resolution spectroscopy (e.g., Liu et al., in prep).

GS~1354$-$64 has been proposed to be a high-inclination system \citep{2009ApJS..181..238C, 2018ApJ...865..134X}, though estimates vary significantly \citep{2023ApJ...951..145L, 2024ApJ...969...40D}. The detection of $4\%$ polarization is broadly consistently with moderate-to-high inclination viewing angles where electron scattering produces stronger net polarization \citep{1977Natur.266..429S, 2009ApJ...701.1175S, 2010ApJ...712..908S}. At lower inclinations, geometric cancellation is expected to suppress the observed PD.

The detection of a $\sim$5~Hz type-C QPO provides further evidence that the source is at the cusp of state transition, possibly with a truncated-disk plus hot inner flow, as commonly thought to be characteristic of an intermediate state. In such geometries, Lense-Thirring precession of the inner flow has been proposed as a mechanism for QPO generation if there is misalignment between the spin axis and the orbital angular momentum of the binary. \citep{2009MNRAS.397L.101I, 2012MNRAS.419.2369I} In principle, this may induce polarization variability detectable through Fourier-based polarimetric timing methods, such as outlined in \cite{2025MNRAS.544.3417E}. Given the intermediate-state of the source during the \ixpe observation, the assumption of stationarity may not be strictly valid. In particular, the QPO frequency is likely evolving. Given the substantial variability observed in the \ixpe light curve of this source, additional stochastic and aperiodic variability of the polarization properties may also be present, associated with, for example, drops in the flux (light curve dips) as investigated in \cite{2022Sci...378..650K} and in \ixpe analysis of the dips in other classes of sources such as the neutron star low-mass X-ray binary GX~13$+$1 \citep{2024A&A...688A.170B, 2025ApJ...979L..47D, 2025ApJ...992...66K, 2026ApJ...997...60R}. Notably, to the extent these are present, they would tend to have a net depolarizing effect.

Taken together, the timing, spectral, and polarimetric properties suggest the following picture: GS~1354$-$64  approached the soft state during its flare but retained a substantial Comptonizing inner flow. The coronal geometry remained radially extended and dynamically important with the resulting X-ray emission dominated by a polarized Compton component, with soft flux and contribution from the enshrouded disk. 
The PD trend can be naturally explained by a radially-extended anisotropic coronal region whose intrinsic polarization efficiency increases with energy, possibly accompanied by a weakly polarized disk. In this framework, the rise in PD reflects the growing dominance and/or higher scattering order of the coronal emission across the \ixpe bandpass, while the constant PA indicates a geometrically coherent symmetry axis for the inner flow. Future radio constraints on the jet position angle would provide important information establishing the orientation of the corona relative to the jet and validate whether GS~1354$-$64 follows the pattern of several BHXBs which have PA aligned with the jet direction. Such constraints will help contextualize the absolute PA measurement presented in this work towards a physically grounded geometric interpretation of the disk–corona configuration. 
%Future observations may further clarify the coronal geometry and disk evolution mechanisms in this system.

\begin{acknowledgments}
This work reports observations obtained with the Imaging X-ray Polarimetry Explorer (IXPE), a joint US (NASA) and Italian (ASI) mission, led by Marshall Space Flight Center (MSFC). The research uses data products provided by the IXPE Science Operations Center (MSFC), using algorithms developed by the IXPE Collaboration (MSFC, Istituto Nazionale di Astrofisica - INAF, Istituto Nazionale di Fisica Nucleare - INFN, ASI Space Science Data Center - SSDC), and distributed by the High-Energy Astrophysics Science Archive Research Center (HEASARC). This research has made use of data from the \nustar mission, a project led by the California Institute of Technology, managed by the Jet Propulsion Laboratory, and funded by the National Aeronautics and Space Administration. Data analysis was performed using the NuSTAR Data Analysis Software (NuSTARDAS), jointly developed by the ASI Science Data Center (SSDC, Italy) and the California Institute of Technology (USA).
This research has made use of MAXI data provided by RIKEN, JAXA and the MAXI team.
The MeerKAT telescope is operated by the South African Radio Astronomy Observatory, which is a facility of the National Research Foundation, an agency of the Department of Science and Innovation.
We acknowledge the use of the ilifu cloud computing facility – www.ilifu.ac.za, a partnership between the University of Cape Town, the University of the Western Cape, Stellenbosch University, Sol Plaatje University and the Cape Peninsula University of Technology. The ilifu facility is supported by contributions from the Inter-University Institute for Data Intensive Astronomy (IDIA – a partnership between the University of Cape Town, the University of Pretoria and the University of the Western Cape), the Computational Biology division at UCT and the Data Intensive Research Initiative of South Africa (DIRISA).

We thank the \ixpe and \nustar SOCs for their coordination for this temperamental source. FC acknowledge financial support by the Istituto Nazionale di Astrofisica (INAF) grant 1.05.24.02.04: ``A multi frequency spectro-polarimetric campaign to explore spin and geometry in Low Mass X-ray Binaries.'' AV acknowledges support from the Academy of Finland grants 355672 and 372881. Nordita is supported in part by NordForsk. MN is a Fonds de Recherche du Quebec – Nature et Technologies (FRQNT) postdoctoral fellow. POP acknowledge financial support from the ``Action Thematiques Phenomenes Extremes et Multimessager" (ATPEM) from the French CNRS and from the French Space Agency CNES. AI acknowledges support by the European Union (ERC, X-MAPS, 101169908). Views and opinions expressed are however those of the author(s) only and do not necessarily reflect those of the European Union or the European Research Council. Neither the European Union nor the granting authority can be held responsible for them. JP and JS acknowledge GACR support from the project 26-226145 and the institutional support from RVO:67985815. The work of LM is funded by the PRIN 2022 - project 2022LWPEXW - ``An X-ray view of compact objects in polarized light'', European Union funding - Next Generation EU, Mission 4 Component 1, CUP C53D23001180006.

\end{acknowledgments}

\facilities{\ixpe \citep{2022JATIS...8b6002W,2021AJ....162..208S}, \maxi \citep{2009PASJ...61..999M}, \nustar \citep{2013ApJ...770..103H}, MeerKAT \citep{2016mks..confE...1J}} %\swift \citep{2005SSRv..120..143B, 2013ApJS..209...14K}

\software{astropy \citep{astropy:2013, astropy:2018, astropy:2022}, NumPy \citep{harris2020array}, SciPy \citep{2020SciPy-NMeth}, IPython \citep{PER-GRA:2007}, bilby \citep{2019ApJS..241...27A}, dynesty \citep{2020MNRAS.493.3132S}, ixpeobssim \citep{2022SoftX..1901194B}, QUEEN-BEE \citep{ravi2025queenbee, 2026ApJ...997...60R}, Stingray \citep{2019ApJ...881...39H, matteo_bachetti_2023_7970570}, SAOImage~DS9 \citep{2003ASPC..295..489J}, XSPEC \citep{1996ASPC..101...17A}}

\bibliography{1354}{}
\bibliographystyle{aasjournalv7}

\end{document}